\documentclass[english]{article}
\usepackage[T1]{fontenc}
\usepackage[latin9]{inputenc}
\usepackage{geometry}
\usepackage{color}
\usepackage{float}
\usepackage{graphicx}
\usepackage{amsmath}
\usepackage{array}
\usepackage{longtable}
\usepackage{amssymb}
\usepackage{babel}
\makeatletter

\providecommand{\tabularnewline}{\\}

\makeatother

\usepackage{babel}
\begin{document}
\title{Efficient quantum circuits for perfect and controlled teleportation
of n-qubit non-maximally entangled states of generalized Bell-type}

\author{Anirban Pathak and Anindita Banerjee  }

\maketitle
\begin{center}
Jaypee Institute of Information Technology University, Noida, India
\par\end{center}

\begin{abstract}
An efficient and economical scheme is proposed for the perfect quantum teleportation of $n$-qubit non-maximally
entangled state of generalized Bell-type. A Bell state is used   as the quantum channel in the proposed scheme.
It is also shown that the controlled teleportation of this $n$-qubit state can be achieved by using a GHZ state
or a GHZ-\emph{like} state as quantum channel. The proposed schemes are economical because for the perfect and
controlled teleportation of $n$-qubit non-maximally entangled state of generalized Bell-type we only need a Bell
state and a tripartite entangled state respectively. It is also established that there exists a family of 12
orthogonal tripartite GHZ-\emph{like} states which can be used as quantum channel for controlled teleportation.
The proposed protocols are critically compared with the existing protocols.

\end{abstract}


\section{Introduction}

 The beauty of quantum circuit lies in the fact that it can perform certain tasks which are impossible in the
classical world. For example, we can consider teleportation and super dense coding. The teleportation is a
quantum task in which an unknown quantum state is transmitted from a sender (Alice) to a spatially separated
receiver (Bob) via an entangled quantum channel and with the help of some classical communications. The original
scheme was proposed by Bennet \emph{et al.} in 1993 \cite{bennettele}. Since then large number of teleportation
schemes and their applications have been reported
\cite{Zhang,Yang,gao,shi1,gor,werner,Karlsson,Li,Lu,Shi2,Vaidman,Cola,joo,baan,Cao}. Some of these teleportation
schemes are also experimentally realized by different groups \cite{qt1,qt2,qt3,qt4,qt5,qt6}. The initial
proposals of teleportation were meant for perfect teleportation of an unknown qubit
$\alpha|0\rangle+\beta|1\rangle$. By perfect teleportation we mean that the success rate of teleportation is
unity. This requires a maximally entangled quantum channel. But immediately after the pioneering work of Bennet
it was realized that teleportation is possible even when the quantum channels are non-maximally entangled. In
that case the success rate of Bob will not be unity and the teleportation scheme would be called probabilistic.

In last two decades several authors have proposed protocols for transformation of more complex quantum states.
For example, schemes for teleportation of EPR pair ($\alpha|00\rangle+\beta|11\rangle)$ \cite{gor}, an arbitrary
two qubit state $a|00\rangle+b|01\rangle+c|10\rangle+d|11\rangle$ \cite{gao} and an arbitrary $n$-qubit state
\cite{Cao} are proposed in recent past. At the same time, possibility of many-party quantum teleportation has
been studied by different groups and these studies lead to schemes for controlled teleportation (CT) or quantum
information splitting (QIS). In these schemes, Alice shares prior entanglement with Bob and at least one Charlie
(supervisor). Charlie is supervisor in the sense that he can control the channel between Alice and Bob. Bob can
properly construct the state sent by Alice if Charlie cooperates (i.e. if Charlie send the value of his
measurements to Bob). Such schemes for controlled teleportation of an unknown qubit are recently studied by
Karlsson \cite{Karlsson}, Jaewoo \cite{joo}, Zhang \emph{et al.} \cite{Zhang} and Yang \emph{et al.} \cite{Yang}
by using different tripartite entangled states as quantum channel. To be precise, Karlsson has used GHZ state,
Zhang \emph{et al.} \cite{Zhang} and Yang \emph{et al.} \cite{Yang} have used GHZ-like state and Joo \cite{joo}
has used W state. Here we would like to note that in 1999, Hillery \cite{Hillery} had proposed a protocol for
quantum secret sharing (QSS). Now if we consider that the quantum secret is a quantum state then a QSS scheme
reduces to a scheme for quantum state sharing (QSTS) or quantum information splitting (QIS). We can visualize
the situation as if a secret (which is a quantum state) is teleported to Charlie and Bob and one of them (Bob)
can obtain the secret (the quantum state) if the other one (Charlie) collaborates. Now it is easy to realize
that in our context, $CT=QIS=QSTS\subset QSS$. Consequently the proposed protocols may found specific
applications in quantum secret communication schemes. In a CT protocol the measurements of Alice and Charlie are
communicated to Bob via classical communication channels and Bob uses these information to choose the unitary
operation to be applied by him. Consequently, any CT protocol may be reduced to usual teleportation scheme
either by keeping the Charlie's bit with Alice or by communicating the bit to Bob. But that would only make the
quantum channel more complex. Further we wish to add that Werner \cite{Werner} had shown that there exist a one
to one correspondence between dense coding and teleportation when these schemes are assumed to be tight. As we
have already mentioned different quantum channels may be used to teleport the same quantum state. So far the
choice of quantum channel is a bit ad hoc and a trend of introducing more and more complex quantum channels have
been observed in recent past. For example, Brown state \cite{brown}, W state \cite{joo}, etc. are recently
introduced as quantum channel for teleportation. But with the present experimental efficiency it is quite
difficult to construct and maintain such multi-particle entangled states. Keeping these facts in mind we have
tried to minimize the quantum channel cost. To do so we have generalized the existing ideas and have shown that
the non-maximally entangled state of generalized Bell-like can be teleported (perfectly or in controlled manner)
by using an optimal quantum channel. The paper also establishes that there exist a family of quantum states of
GHZ-type which can be used for perfect and probabilistic controlled teleportation. The paper also produces some
of the recently reported results as special cases of the more general results obtained here. For example,
teleportation schemes proposed by Cola \cite{Cola}, An Ba \cite{baan}, Zhang \emph{et al.} \cite{Zhang} and Yang
\emph{et al.} \cite{Yang} etc. can be obtained as special cases of the present work.

In section 2 we have proposed a new scheme for teleportation of an unknown $n$-qubit non-maximally entangled
state of the form $\alpha\left|x\right\rangle +\beta\left|\bar{x}\right\rangle $ by using a Bell state. In
section 3 we have shown that the CT of the same state is possible if we use a GHZ state as quantum channel. In
section 4 we have shown that the CT scheme described in section 3 may also be achieved by using any member of a
family of 12 orthogonal GHZ\emph{-like} states as quantum channel. Some of the existing results are also
obtained as special cases of this protocol. Finally section 5 is dedicated to conclusions.


\section{Protocol for teleportation of an unknown $n$-qubit non-maximally entangled
state of the form  $\alpha\left|x\right\rangle \pm\beta\left|\bar{x}\right\rangle $ using Bell state as quantum
channel}

Suppose we wish to teleport an $n$-qubit non-maximally entangled state of the form

\begin{equation}
\begin{array}{cc}
\left|\psi^{\pm}\right\rangle = & \alpha\left|x\right\rangle \pm\beta\left|\bar{x}\right\rangle
\end{array}\label{eq:t1}\end{equation} where $|\alpha|^{2}+|\beta|^{2}=1$, $x$ varies from $0$ to $2^{n}-1$ and
$\bar{x}=1^{\otimes n}\oplus x$ in modulo 2 arithmetic. This state will reduce to generalized Bell state (GBS)
\cite{the:AIP-Manu} for $\alpha=\beta=\frac{1}{\sqrt{2}}$. Usual Bell state and GHZ state are special cases of
GBS for $n=2$ and $n=3$ respectively. Consequently a successful teleportation scheme for the above state will be
able to teleport GHZ state, Bell state in general and EPR states in particular. Since
$\frac{1}{\sqrt{2}}\left(\left|x\right\rangle \pm\beta\left|\bar{x}\right\rangle \right)$ is called generalized
Bell state \cite{the:AIP-Manu}, we may call $\alpha\left|x\right\rangle +\beta\left|\bar{x}\right\rangle $ as
generalized Bell-type state. Now, without loss of generality we may consider the unknown generalized Bell-type
state to be teleported as

\begin{equation}
\begin{array}{ccc}
\left|\psi\right\rangle _{unknown} & = & \alpha\left|x\right\rangle +\beta\left|\bar{x}\right\rangle
=\alpha\left|x_{1}x_{2}..x_{n}\right\rangle +\beta\left|\bar{x}_{1}\bar{x}_{2}..\bar{x}_{n}\right\rangle
\end{array}\label{eq:state}\end{equation} where $x_{i}\in\left\{ 0,1\right\} $. Alice may teleport this unknown
state to bob in following five steps:

Step1: Let Alice and Bob share a Bell state of the form \begin{equation}
|\psi\rangle\ensuremath{_{channel}}=\frac{1}{\sqrt{2}}\left(\left|00\right\rangle +\left|11\right\rangle
\right)_{AB}.\label{eq:t2}\end{equation} The first qubit of this maximally entangled state is kept with Alice
and the second qubit is sent to Bob. Thus this particular Bell state (\ref{eq:t2}) constitutes our quantum
channel. The proposed protocol is valid for any Bell state. Now the input state of the circuit (i.e. the state
of the system after Alice receives the unknown $n$-qubit state to be teleported) is

\begin{equation}
\begin{array}{lcl}
\left|\Psi_{in}\right\rangle  & = & \left|\psi\right\rangle _{unknown}\otimes|\psi\rangle\ensuremath{_{channel}}\\
 & = & \frac{1}{\sqrt{2}}\left[\alpha\left|x_{1}x_{2}x_{3}..x_{n}00\right\rangle +\alpha\left|x_{1}x_{2}x_{3}..x_{n}11\right\rangle +\beta\left|\bar{x_{1}}\bar{x_{2}}\bar{x_{3}..}\bar{x_{n}}00\right\rangle +\beta\left|\bar{x_{1}}\bar{x_{2}}\bar{x_{3}..}\bar{x_{n}}11\right\rangle \right]_{A_{1}A_{2}..A_{n}AB}\end{array}\label{eq:t3}\end{equation}

\begin{figure}
\centering{}\includegraphics[scale=0.7]{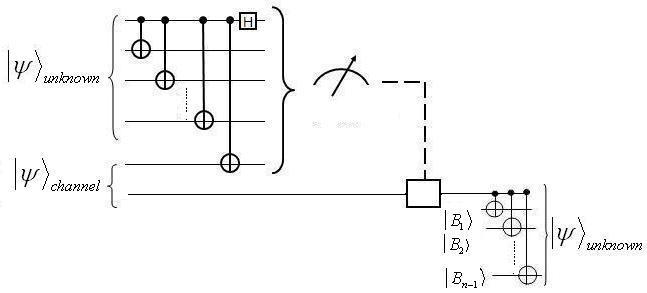} \caption{Quantum circuit for efficient teleportation of
unknown $n$-qubit non-maximally entangled state of the form $\alpha\left|x\right\rangle
+\beta\left|\bar{x}\right\rangle $ by using a Bell state as quantum channel.}
\end{figure}

Here the first $(n+1)$ qubits are with Alice and the last qubit is with Bob.
Step2: Alice performs $n$ Cnot
operations on her qubits (as shown in Fig. 1) by using the first qubit ($A_{1}$) as the control qubit and the
next $n$ qubits as target qubits. This operations transform the input state of the system (\ref{eq:t3}) to
\begin{equation}
\begin{array}{lcl}
\left|\Psi\right\rangle _{1} & = & \frac{1}{\sqrt{2}}\left[\alpha\left|x_{1}\left(x_{1}\oplus x_{2}\right)\left(x_{1}\oplus x_{3}\right)...\left(x_{1}\oplus x_{n}\right)\left(x_{1}\oplus0\right)0\right\rangle \right.\\
 & + & \alpha\left|x_{1}\left(x_{1}\oplus x_{2}\right)\left(x_{1}\oplus x_{3}\right)...\left(x_{1}\oplus x_{n}\right)\left(x_{1}\oplus1\right)1\right\rangle \\
 & + & \beta\left|\bar{x_{1}}\left(\bar{x_{1}}\oplus\bar{x_{2}}\right)\left(\bar{x_{1}}\oplus\bar{x_{3}}\right)...\left(\bar{x_{1}}\oplus\bar{x_{n}}\right)\left(\bar{x_{1}}\oplus0\right)0\right\rangle \\
 & + & \left.\beta\left|\bar{x_{1}}\left(\bar{x_{1}}\oplus\bar{x_{2}}\right)\left(\bar{x_{1}}\oplus\bar{x_{3}}\right)...\left(\bar{x_{1}}\oplus\bar{x_{n}}\right)\left(\bar{x_{1}}\oplus1\right)1\right\rangle \right]_{A_{1}A_{2}..A_{n}AB}\end{array}.\label{eq:4}\end{equation}
 Here we may use the identities \begin{equation}
\begin{array}{lcl}
\left(a\oplus b\right) & = & \left(\bar{a}\oplus\bar{b}\right)\\
\left(a\oplus0\right) & = & a\\
\left(a\oplus1\right) & = & \bar{a}\end{array}\label{eq:5}\end{equation} where $\left\{ a,b\right\} \in\left\{
0,1\right\} $, to obtain \begin{equation}
\begin{array}{lcl}
\left|\Psi\right\rangle _{1} & = & \frac{1}{\sqrt{2}}\left[\alpha\left|x_{1}\left(x_{1}\oplus x_{2}\right)\left(x_{1}\oplus x_{3}\right)...\left(x_{1}\oplus x_{n}\right)x_{1}0\right\rangle \right.\\
 & + & \alpha\left|x_{1}\left(x_{1}\oplus x_{2}\right)\left(x_{1}\oplus x_{3}\right)...\left(x_{1}\oplus x_{n}\right)\bar{x_{1}}1\right\rangle \\
 & + & \beta\left|\bar{x_{1}}\left(x_{1}\oplus x_{2}\right)\left(x_{1}\oplus x_{3}\right)....\left(\bar{x_{1}}\oplus\bar{x_{n}}\right)\bar{x_{1}}0\right\rangle \\
 & + & \left.\beta\left|\bar{x_{1}}\left(x_{1}\oplus x_{2}\right)\left(x_{1}\oplus x_{3}\right)....\left(\bar{x_{1}}\oplus\bar{x_{n}}\right)x_{1}1\right\rangle \right]_{A_{1}A_{2}..A_{n}AB}\end{array}.\label{eq:t5}\end{equation}

Step 3: Alice applies a Hadamard operation%
\footnote{A Hadamard operation is defined by

$H\left|a\right\rangle =\frac{\left(-1\right)^{a}\left|a\right\rangle +\left|\bar{a}\right\rangle }{\sqrt{2}}$
where $a\in\{0,1\}$%
} on her first qubit $\left|x_{1}\right\rangle $ to transform the state $|\Psi\rangle_{1}$ to

\begin{equation}
\begin{array}{lcl}
\left|\Psi\right\rangle _{2} & = & \frac{1}{2}\left[\alpha(-1)^{x_{1}}\left|x_{1}\left(x_{1}\oplus x_{2}\right)\left(x_{1}\oplus x_{3}\right)...\left(x_{1}\oplus x_{n}\right)x_{1}0\right\rangle \right.\\
 & + & \alpha\left|\bar{x_{1}}\left(x_{1}\oplus x_{2}\right)\left(x_{1}\oplus x_{3}\right)...\left(x_{1}\oplus x_{n}\right)x_{1}0\right\rangle \\
 & + & \alpha(-1)^{x_{1}}\left|x_{1}\left(x_{1}\oplus x_{2}\right)\left(x_{1}\oplus x_{3}\right)...\left(x_{1}\oplus x_{n}\right)\bar{x_{1}}1\right\rangle \\
 & + & \alpha\left|\bar{x_{1}}\left(x_{1}\oplus x_{2}\right)\left(x_{1}\oplus x_{3}\right)...\left(x_{1}\oplus x_{n}\right)\bar{x_{1}}1\right\rangle \\
 & + & \beta(-1)^{\bar{x_{1}}}\left|\bar{x_{1}}\left(x_{1}\oplus x_{2}\right)\left(x_{1}\oplus x_{3}\right)...\left(x_{1}\oplus x_{n}\right)\bar{x_{1}}0\right\rangle \\
 & + & \beta\left|x_{1}\left(x_{1}\oplus x_{2}\right)\left(x_{1}\oplus x_{3}\right)...\left(x_{1}\oplus x_{n}\right)\bar{x_{1}}0\right\rangle \\
 & + & \beta(-1)^{\bar{x_{1}}}\left|\bar{x_{1}}\left(x_{1}\oplus x_{2}\right)\left(x_{1}\oplus x_{3}\right)...\left(x_{1}\oplus x_{n}\right)x_{1}1\right\rangle \\
 & + & \left.\beta\left|x_{1}\left(x_{1}\oplus x_{2}\right)\left(x_{1}\oplus x_{3}\right)...\left(x_{1}\oplus x_{n}\right)x_{1}1\right\rangle \right]_{A_{1}A_{2}..A_{n}AB}\end{array}.\label{eq:t6}\end{equation}
By using $\left(x_{1}\oplus x_{n}\right)=e_{n-1}$ we can obtain \begin{equation}
\begin{array}{lcl}
\left|\Psi\right\rangle _{2} & = & \frac{1}{2}\left[\alpha(-1)^{x_{1}}\left|x_{1}e_{1}e_{2}...e_{n-1}x_{1}0\right\rangle \right.\\
 & + & \alpha\left|\bar{x_{1}}e_{1}e_{2}...e_{n-1}x_{1}0\right\rangle \\
 & + & \alpha(-1)^{x_{1}}\left|x_{1}e_{1}e_{2}...e_{n-1}\bar{x_{1}}1\right\rangle \\
 & + & \alpha\left|\bar{x_{1}}e_{1}e_{2}...e_{n-1}\bar{x_{1}}1\right\rangle \\
 & + & \beta(-1)^{\bar{x_{1}}}\left|\bar{x_{1}}e_{1}e_{2}...e_{n-1}\bar{x_{1}}0\right\rangle \\
 & + & \beta\left|x_{1}e_{1}e_{2}...e_{n-1}\bar{x_{1}}0\right\rangle \\
 & + & \beta(-1)^{\bar{x_{1}}}\left|\bar{x_{1}}e_{1}e_{2}...e_{n-1}x_{1}1\right\rangle \\
 & + & \left.\beta\left|x_{1}e_{1}e_{2}...e_{n-1}x_{1}1\right\rangle \right]_{A_{1}A_{2}..A_{n}AB}\end{array}.\label{eq:t8}\end{equation}
Here we can easily observe that Alice's first and last qubits $(A_{1}$ and $A)$ are entangled with the Bob's
qubit $(B)$ and the remaining $(n-1)$ qubits of Alice (i.e. $A_{2}A_{3}..A_{n}$) are separable.

Step 4: Alice measures all her qubits in computational basis and classically communicate the results
 to Bob. For simplicity of visualization we may consider this measurement as two step
process in which Alice measures the separable qubits first by applying a measurement
$M_{n-1}=M_{A_{2}A_{3}..A_{n}}$ in computational basis. This measurement reduces the state to \begin{equation}
\begin{array}{lcl}
\left|\Psi\right\rangle _{3} & = & \frac{1}{2}\left[\left(-1\right)^{x_{1}}\left|x_{1}x_{1}\right\rangle _{A_{1}A}\left(\alpha\left|0\right\rangle +\left(-1\right)^{x_{1}}\beta\left|1\right\rangle \right)_{B}\right.\\
 & + & \left|\bar{x_{1}}x_{1}\right\rangle _{A_{1}A}\left(\alpha\left|0\right\rangle +\left(-1\right)^{\bar{x_{1}}}\beta\left|1\right\rangle \right)_{B}\\
 & + & \left(-1\right)^{x_{1}}\left|x_{1}\bar{x_{1}}\right\rangle _{A_{1}A}\left(\alpha\left|1\right\rangle +\left(-1\right)^{x_{1}}\beta\left|0\right\rangle \right)_{B}\\
 & + & \left|\bar{x_{1}}\bar{x_{1}}\right\rangle _{A_{1}A}\left.\left(\alpha\left|1\right\rangle +\left(-1\right)^{\bar{x_{1}}}\beta\left|0\right\rangle \right)_{B}\right]\end{array}.\label{eq:t10}\end{equation}
Then Alice measures her entangled qubits by applying a measurement $M_{A_{1}A}$ . We will show in the next step
that Bob can reconstruct the initial state (\ref{eq:state}) by using the values of $e_{1}..e_{n-1}$provided he
has $\alpha\left|x_{1}\right\rangle +\beta\left|\bar{x_{1}}\right\rangle $.

Step 5: Now it is easy to observe from (\ref{eq:t10}) that Bob may construct $\alpha\left|x_{1}\right\rangle
+\beta\left|\bar{x_{1}}\right\rangle $ by using the result of $M_{A_{1}A}$. This is so because from from
(\ref{eq:t10}) we observe that if the measurement on Alice's first qubit yield 0 then we do not need to change
the relative phase of Bob's state. On the other hand if the measurement yield 1 then Bob has to apply a phase
flip operation {}``Z'' to get the correct relative phase. Again if the second measurement yield zero then the
bit values are proper but if the second measurement yield 1 then Bob has to apply a bit flip operation {}``X''
to obtain $\alpha\left|x_{1}\right\rangle +\beta\left|\bar{x_{1}}\right\rangle $. To be more precise, according
to the outcome of $M_{A_{1}A}$, Bob
applies suitable gates as shown in Table 1 to his qubit%
\footnote{global phase is ignored here.%
}. After constructing $\alpha\left|x_{1}\right\rangle +\beta\left|\bar{x}_{1}\right\rangle $, Bob creates
$(n-1)$ ancilla qubits $\left|B_{1}B_{2}..B_{n-1}\right\rangle $ in accordance to the outcome of $M_{n-1}$ such
that $\left|B_{1}B_{2}..B_{n-1}\right\rangle =\left|e_{1}..e_{n-1}\right\rangle $. Then Bob performs $(n-1)$
Cnot operations on his qubits by using the first qubit as control qubit and rest $(n-1$) qubits as target
qubits. and retrieve the unknown state (\ref{eq:state}). As the measurement and classical communication can be
replaced by non-local unitary operation we can convert the generalized circuit shown in Fig. 1 to a circuit for
measurement less teleportation of a 2-qubit non-maximally entangled quantum state as shown in Fig 2. This is
done only for comparison purpose. For teleporting a 2-qubit non-maximally entangled quantum state we need a Bell
state as quantum channel, 9 gates and quantum cost \cite{qc} of our circuit is 7 where as Gorbachev's circuit
\cite{gor} for the same purpose requires a GHZ channel, 16 gates and quantum cost of the circuit is 10. Thus the
special case of our circuit (shown in Fig. 2) has lesser gate count, lesser quantum cost and lesser quantum
channel cost as compared to Gorbachev's circuit \cite{gor} for the same purpose.

\begin{table}
\begin{centering}
{\small }\begin{tabular}{|c|c|c|c|c|c|} \hline {\small $M_{A_{1}A}$} & {\small Bob's state } & {\small Gates } &
{\small Bob's state after production of Ancilla} & {\small Bob's final state }\tabularnewline \hline \hline
{\small $\left|00\right\rangle $} & {\small $\alpha\left|0\right\rangle +\beta\left|1\right\rangle $ } & {\small
I}  & {\small $\alpha\left|0e_{1}e_{2}..e_{n-1}\right\rangle
+\beta\left|1\bar{e_{1}}\bar{e_{2}}..\bar{e_{n-1}}\right\rangle $ } & {\small
$\alpha\left|x_{1}...x_{n}\right\rangle +\beta\left|\bar{x_{1}}...\bar{x_{n}}\right\rangle $ }\tabularnewline
\hline {\small $\left|01\right\rangle $} & {\small $\alpha\left|1\right\rangle +\beta\left|0\right\rangle $ } &
{\small X}  & {\small $\alpha\left|0e_{1}e_{2}..e_{n-1}\right\rangle
+\beta\left|1\bar{e_{1}}\bar{e_{2}}..\bar{e_{n-1}}\right\rangle $ } & {\small
$\alpha\left|x_{1}...x_{n}\right\rangle +\beta\left|\bar{x_{1}}...\bar{x_{n}}\right\rangle $}\tabularnewline
\hline {\small $\left|10\right\rangle $} & {\small $\alpha\left|0\right\rangle -\beta\left|1\right\rangle $ } &
{\small Z}  & {\small $\alpha\left|0e_{1}e_{2}..e_{n-1}\right\rangle
-\beta\left|1\bar{e_{1}}\bar{e_{2}}..\bar{e_{n-1}}\right\rangle $ } & {\small
$\alpha\left|x_{1}...x_{n}\right\rangle +\beta\left|\bar{x_{1}}...\bar{x_{n}}\right\rangle $}\tabularnewline
\hline {\small $\left|11\right\rangle $} & {\small $\alpha\left|1\right\rangle -\beta\left|0\right\rangle $ } &
{\small iY}  & {\small $\alpha\left|0e_{1}e_{2}..e_{n-1}\right\rangle
-\beta\left|1\bar{e_{1}}\bar{e_{2}}..\bar{e_{n-1}}\right\rangle $ } & {\small
$\alpha\left|x_{1}...x_{n}\right\rangle +\beta\left|\bar{x_{1}}...\bar{x_{n}}\right\rangle $}\tabularnewline
\hline
\end{tabular}
\par\end{centering}{\small \par}

\caption{Quantum gates applied by Bob according to different measurement outcomes of Alice.}
\end{table}

\begin{figure}
\centering{} \centering{}\includegraphics[scale=0.7]{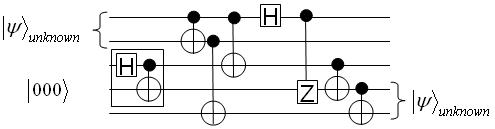}
 \caption{Measurement less quantum circuit
for efficient teleportation of  $\alpha\left|x\right\rangle +\beta\left|\bar{x}\right\rangle $ by using a Bell
state.}

\end{figure}

\section{Controlled teleportation of an unknown $n$-qubit non-maximally entangled state
of the form\textmd{\large{} }\textmd{\normalsize $\alpha\left|x\right\rangle \pm\beta\left|\bar{x}\right\rangle
$ }using a GHZ state as quantum channel{\normalsize .}}

It is already mentioned in the introduction that the CT is equivalent to QIS and QSTS. In this multiparty
teleportation scheme Alice sends a secret message to a party say Bob in cooperation with another party Charlie.
Bob can create the secret message with the information obtained from Alice and Charlie. In our case the message
is  generalized $n$-qubit state of the form (\ref{eq:state}).

Let Alice, Bob and Charlie share a GHZ state as quantum channel \begin{equation} \left|\psi\right\rangle
_{ABC}=\frac{1}{\sqrt{2}}\left(\left|000\right\rangle +\left|111\right\rangle
\right)_{ABC}\label{eq:mt2}\end{equation}

The first qubit of the channel is kept with Alice and the second and third qubits are sent to Bob and Charlie
respectively.  The input state of this circuit for CT is shown in Fig. 3.
\begin{equation}
\begin{array}{lcl}
\left|\Psi\right\rangle  & = & |\psi\rangle_{unknown}\otimes|\psi\rangle\ensuremath{_{channel}}\\
 & = & \left[\alpha\left|x_{1}x_{2}x_{3}..x_{n}000\right\rangle +\alpha\left|x_{1}x_{2}x_{3}..x_{n}111\right\rangle \right.\\
 & + & \left.\beta\left|\bar{x_{1}}\bar{x_{2}}\bar{x_{3}..}\bar{x_{n}}000\right\rangle +\beta\left|\bar{x_{1}}\bar{x_{2}}\bar{x_{3}..}\bar{x_{n}}111\right\rangle \right]_{A_{1}A_{2}..A_{n}ABC}\end{array}.\label{eq:mt3}\end{equation}

\begin{figure}
\centering{}\textbackslash{}\includegraphics[scale=0.7]{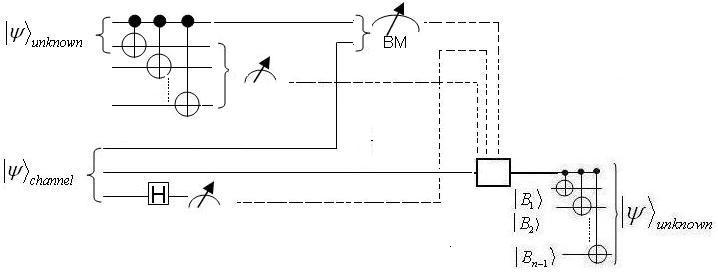}\caption{Circuit for controlled
teleportation of $\alpha\left|x\right\rangle +\beta\left|\bar{x}\right\rangle $ using GHZ state. }
\end{figure}
After application of $(n-1)$ Cnots, the input state of the system transforms to \begin{equation}
\begin{array}{lcl}
\left|\Psi\right\rangle _{1} & = & \left[\left|\alpha x_{1}\left(x_{1}\oplus x_{2}\right)..\left(x_{1}\oplus x_{n}\right)000\right\rangle \right.\\
 & + & \left|\alpha x_{1}\left(x_{1}\oplus x_{2}\right)..\left(x_{1}\oplus x_{n}\right)\left(x_{1}\oplus1\right)111\right\rangle \\
 & + & \left|\beta\bar{x_{1}}\left(\bar{x_{1}}\oplus\bar{x_{2}}\right)..\left(\bar{x_{1}}\oplus\bar{x_{n}}\right)000\right\rangle \\
 & + & \left.\left|\bar{\beta x_{1}}\left(\bar{x_{1}}\oplus\bar{x_{2}}\right)..\left(\bar{x_{1}}\oplus\bar{x_{n}}\right)111\right\rangle \right]_{A_{1}A_{2}..A_{n}ABC}\end{array}.\label{eq:mt4}\end{equation}
Upto this point the CT scheme is similar to the usual teleportation scheme described in last section but after
this point Charlie applies the Hadamard transformation on her qubit. This operation of Charlie transforms the
state of the system to

\begin{equation}
\begin{array}{lcl}
\left|\Psi\right\rangle _{2} & = & \frac{1}{2}\left[\alpha\left|x_{1}\left(x_{1}\oplus x_{2}\right)\left(x_{1}\oplus x_{3}\right)...\left(x_{1}\oplus x_{n}\right)000\right\rangle \right.\\
 & + & \alpha\left|x_{1}\left(x_{1}\oplus x_{2}\right)\left(x_{1}\oplus x_{3}\right)...\left(x_{1}\oplus x_{n}\right)001\right\rangle \\
 & + & \alpha\left|x_{1}\left(x_{1}\oplus x_{2}\right)\left(x_{1}\oplus x_{3}\right)...\left(x_{1}\oplus x_{n}\right)110\right\rangle \\
 & - & \alpha\left|x_{1}\left(x_{1}\oplus x_{2}\right)\left(x_{1}\oplus x_{3}\right)...\left(x_{1}\oplus x_{n}\right)111\right\rangle \\
 & + & \beta\left|\bar{x_{1}}\left(x_{1}\oplus x_{2}\right)\left(x_{1}\oplus x_{3}\right)...\left(x_{1}\oplus x_{n}\right)000\right\rangle \\
 & + & \beta\left|\bar{x_{1}}\left(x_{1}\oplus x_{2}\right)\left(x_{1}\oplus x_{3}\right)...\left(x_{1}\oplus x_{n}\right)001\right\rangle \\
 & + & \beta\left|\bar{x_{1}}\left(x_{1}\oplus x_{2}\right)\left(x_{1}\oplus x_{3}\right)...\left(x_{1}\oplus x_{n}\right)110\right\rangle \\
 & - & \left.\beta\left|\bar{x_{1}}\left(x_{1}\oplus x_{2}\right)\left(x_{1}\oplus x_{3}\right)...\left(x_{1}\oplus x_{n}\right)111\right\rangle \right]_{A_{1}A_{2}..A_{n}ABC}\end{array}\label{eq:t10-1}\end{equation}
At this point Alice applies $M_{n-1}$(i.e she measures the qubits $A_{1}A_{2}...A_{n}$ in computational basis).
This measurement reduces the state of the system to \begin{equation}
|S\rangle_{A_{1}AB}|0\rangle_{C}+|T\rangle_{A_{1}AB}|1\rangle_{C}\label{eq:stateprime}\end{equation} where
$|S\rangle_{A_{1}AB}$ and $|T\rangle_{A_{1}AB}$ can be expanded as\begin{equation}
\begin{array}{l}
\begin{array}{lcl}
|S\rangle_{A_{1}AB} & = & \frac{1}{4}\left(\alpha\left|x_{1}00\right\rangle +\alpha\left|x_{1}11\right\rangle +\beta\left|\bar{x_{1}}00\right\rangle +\beta\left|\bar{x_{1}11}\right\rangle \right)_{A_{1}AB}\\
 & = & \frac{1}{2\sqrt{2}}\left(\left(\frac{\left|x_{1}0\right\rangle +\left|\bar{x_{1}}1\right\rangle }{\sqrt{2}}\right)_{A_{1}A}\otimes(\alpha\left|0\right\rangle +\beta|1)_{B}+\left(\frac{\left|x_{1}0\right\rangle -\left|\bar{x_{1}}1\right\rangle }{\sqrt{2}}\right)_{A_{1}A}\otimes(\alpha\left|0\right\rangle -\beta\left|1\right\rangle )_{B}\right.\\
 & + & \left.\left(\frac{\left|x_{1}1\right\rangle +\left|\bar{x_{1}}0\right\rangle }{\sqrt{2}}\right)_{A_{1}A}\otimes(\alpha\left|1\right\rangle +\beta\left|0\right\rangle )_{B}+\left(\frac{\left|x_{1}1\right\rangle -\left|\bar{x_{1}}0\right\rangle }{\sqrt{2}}\right)_{A_{1}A}\otimes(\alpha\left|1\right\rangle -\beta\left|0\right\rangle )_{B}\right)\end{array}\end{array}\label{eq:t102}\end{equation}
 and

\begin{table}
\begin{centering}
\begin{tabular}{|c|c|c|c|c|c|}
\hline $M_{C}$ & $M_{A_{1}A}$ & Gate $U_{B}$ & $M_{C}$ & $M_{A_{1}A}$ & Gate $U_{B}$\tabularnewline \hline
$\left|0\right\rangle $ & $\psi_{12}^{+}$  & I & $\left|1\right\rangle $ & $\psi_{12}^{+}$  & Z\tabularnewline
\hline $\left|0\right\rangle $ & $\psi_{12}^{-}$  & Z & $\left|1\right\rangle $ & $\psi_{12}^{-}$  &
I\tabularnewline \hline $\left|0\right\rangle $ & $\phi_{12}^{+}$  & X & $\left|1\right\rangle $ &
$\phi_{12}^{+}$  & iY\tabularnewline \hline $\left|0\right\rangle $ & $\phi_{12}^{-}$  & iY &
$\left|1\right\rangle $ & $\phi_{12}^{-}$  & X\tabularnewline \hline
\end{tabular}
\par\end{centering}

\caption{Table for CT using GHZ as quantum channel.}

\end{table}

\begin{equation}
\begin{array}{l}
\begin{array}{lcl}
|T\rangle_{A_{1}AB} & = & \frac{1}{4}\left(\alpha\left|x_{1}00\right\rangle -\alpha\left|x_{1}11\right\rangle +\beta\left|\bar{x_{1}}00\right\rangle -\beta\left|\bar{x_{1}}11\right\rangle \right)\\
 & = & \frac{1}{2\sqrt{2}}\left(\left(\frac{\left|x_{1}0\right\rangle +\left|\bar{x_{1}}1\right\rangle }{\sqrt{2}}\right)_{A_{1}A}\otimes(\alpha\left|0\right\rangle -\beta|1)_{B}+\left(\frac{\left|x_{1}0\right\rangle -\left|\bar{x_{1}}1\right\rangle }{\sqrt{2}}\right)_{A_{1}A}\otimes(\alpha\left|0\right\rangle +\beta\left|1\right\rangle )_{B}\right.\\
 & - & \left.\left(\frac{\left|x_{1}1\right\rangle +\left|\bar{x_{1}}0\right\rangle }{\sqrt{2}}\right)_{A_{1}A}\otimes(\alpha\left|1\right\rangle -\beta\left|0\right\rangle )_{B}-\left(\frac{\left|x_{1}1\right\rangle -\left|\bar{x_{1}}0\right\rangle }{\sqrt{2}}\right)_{A_{1}A}\otimes(\alpha\left|1\right\rangle +\beta\left|0\right\rangle )_{B}\right)\end{array}\end{array}\label{: eq:t103-1}\end{equation}
respectively. From (\ref{eq:t102}) and (\ref{: eq:t103-1}) one can easily see that if Alice measures $A_{1}A$
using Bell analyzers $M_{BM}$ while Charlie measures his qubit by a measurement $M_{C}$ in computational basis
then the state of Bob reduces to a one qubit state which can be transformed into the state
$\alpha|x_{1}\rangle+\beta|\bar{x}_{1}\rangle$ by applying an unitary transformation $U_{B}$ as shown in Table
2. The choice of the particular unitary operation depends on the outcome of the measurement of Alice ($M_{BM})$
and that of Charlie ($M_{C}$). Therefore, it is required that the Alice and Charlie sent their measurement
outcomes $M_{BM}$ and $M_{C}$  to Bob using classical channels and after receiving them, Bob chooses the unitary
operations as per the Table 2. According to measurement $M_{n-1}$ Bob will create n-1 ancillas
$\left|B_{1}B_{2}..B_{n-1}\right\rangle $, such that $\left|B_{1}B_{2}..B_{n-1}\right\rangle $=
$\left|e_{1}..e_{n-1}\right\rangle $. Finally Bob performs $n-1$ Cnot operations as shown in Fig. 3 to
reconstruct the unknown state (\ref{eq:state}).


\section{Controlled teleportation of an unknown n-qubit non-maximally entangled
state of the form\textmd{\normalsize{} $\alpha\left|x\right\rangle +\beta\left|\bar{x}\right\rangle $} using
GHZ-\emph{like} state as quantum channel}

Recently it is reported by Zhang \emph{et al.} \cite{Zhang} that CT of an unknown qubit is possible if Alice,
Bob and Charlie share a GHZ-\emph{like} tripartite maximally entangled state of the from
$|\phi\rangle=\frac{1}{2}\left(|000\rangle+|110\rangle+|101\rangle+|011\rangle\right)$. Almost in the same time
Yang \emph{et al.} have independently shown that the same task may be achieved by using
$|\phi\rangle=\frac{1}{2}\left(|001\rangle+|100\rangle+|010\rangle+|111\rangle\right)$.

Here we that have shown that these states are not unique and there exists an underlying symmetry. To be precise,
not only CT of an unknown qubit but CT of an $n$-qubit non-maximally entangled state of the form
(\ref{eq:state}) is possible if we use quantum channel of the form \begin{equation}
\frac{(\psi_{i}\left|0\right\rangle +\psi_{j}\left|1\right\rangle )}{\sqrt{2}}\label{eq:a}\end{equation}
 where $i,j\epsilon\left\{ 0,1,2,3\right\} $, $i\neq j$ and $\psi_{i,j}$
are Bell states usually denoted as \begin{equation}
\begin{array}{ccc}
\psi_{0} & =\psi_{00}= & \psi^{+}=\frac{\left|00\right\rangle +\left|11\right\rangle }{\sqrt{2}}\\
\psi_{1} & =\psi_{01}= & \phi^{+}=\frac{\left|01\right\rangle +\left|10\right\rangle }{\sqrt{2}}\\
\psi_{2} & =\psi_{10}= & \psi^{-}=\frac{\left|00\right\rangle -\left|11\right\rangle }{\sqrt{2}}\\
\psi_{3} & =\psi_{11}= & \phi^{-}=\frac{\left|01\right\rangle -\left|10\right\rangle
}{\sqrt{2}}\end{array}.\label{eq:bell1}\end{equation}
 For the calculational simplicity we may use a more compact notation
in which a Bell state is written as \begin{equation}
\begin{array}{ccc}
\psi_{i} & = & \frac{\left|0,y\right\rangle +\left(-1^{x}\right)\left|1,\overline{y}\right\rangle
}{\sqrt{2}}\end{array}\label{eq:bell-compact}\end{equation}
 where $x,y$ are two digit binary representation of $i$ and $j$
respectively for example for $\psi_{2}=\psi_{10}$, $x=1$ and $y=0$. Now we can use this compact notation
(\ref{eq:bell-compact}) to describe the state of the quantum channel in general as \begin{equation}
\begin{array}{cl}
\frac{1}{\sqrt{2}}\left(\psi_{i}\left|0\right\rangle +\psi_{j}\left|1\right\rangle \right)_{ABC} & =\frac{1}{\sqrt{2}}\left(\psi_{x,y}\left|0\right\rangle +\psi_{x',y'}\left|1\right\rangle \right)_{ABC}\\
 & =\frac{1}{2}\left(|0y0\rangle+(-1)^{x}|1\bar{y}0\rangle+|0y^{\prime}1\rangle+(-1)^{x}|1\bar{y^{\prime}}1\rangle\right)_{ABC}\end{array}.\label{eq:channel2}\end{equation}

\begin{figure}
\centering{}\includegraphics[scale=0.7]{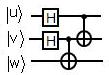}\caption{A circuit to create GHZ-\emph{like}
states$\left|u\right\rangle
$$\left|v\right\rangle \left|w\right\rangle $ are the input bits to the circuit and the output is a
GHZ-\emph{like} state}
\end{figure}

The subscripts $ABC$ used in (\ref{eq:channel2}) indicates that the first qubit of the channel is kept with
Alice and the second and third qubits are sent to Bob and Charlie respectively. The quantum circuit for
generation of these states is given in Fig 4. Now we may perform CT of an $n$-qubit state of the form
(\ref{eq:state}) by using (\ref{eq:channel2}) as quantum channel. The circuit designed for this purpose is shown
in Fig. 5.

\begin{figure}
\centering{}\includegraphics[scale=0.6]{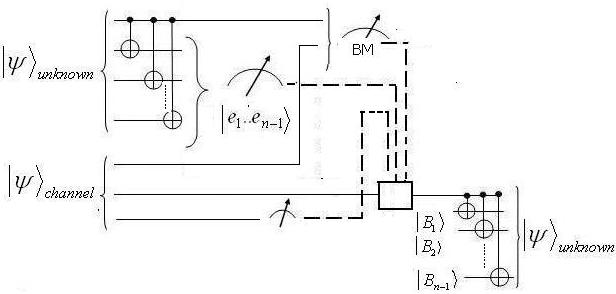}\caption{CT of $\alpha\left|x\right\rangle
+\beta\left|\bar{x}\right\rangle $ using GHZ -\emph{like} quantum state as quantum channel..}

\end{figure}

Here the unknown state to be teleported is (\ref{eq:state}) and the quantum channel to be used for the purpose
is (\ref{eq:channel2}). Consequently the input state of the system is
\begin{equation}
\begin{array}{lcl}
|\Psi\rangle_{1} & = & |\psi\rangle\ensuremath{_{unknown}}\otimes|\psi\rangle\ensuremath{_{channel}}\\
 & = & \left[\left(\alpha\left|a_{1}a_{2}..a_{n}\right\rangle +\beta\left|\bar{a_{1}}\bar{a_{2}}..\bar{a_{n}}\right\rangle \right)\otimes\right.\\
 &  & \left.\frac{1}{2}\left(|0y0\rangle+(-1)^{x}|1\bar{y}0\rangle+|0y^{\prime}1\rangle+(-1)^{x}|1\bar{y^{\prime}}1\rangle\right)\right]_{A_{1}A_{2}..A_{n}ABC}.\end{array}\label{eq:ac}\end{equation}

Now Alice performs $n$-Cnot operations on her qubits (this step is same as step 2 of Section 2) in accordance
with Fig. 5. This operation transforms the input state $|\Psi\rangle_{1}$ into

\[
\begin{array}{lcl}
|\Psi\rangle_{2} & = & \left(\alpha\left|a_{1}\left(a_{1}\oplus a_{2}\right)...\left(a_{1}\oplus a_{n}\right)\right\rangle +\beta\left|\bar{a_{1}}\left(\bar{a_{1}}\oplus\bar{a_{2}}\right)...\left(\bar{a_{1}}\oplus\bar{a_{n}}\right)\right\rangle \right)\otimes\\
 &  & \frac{1}{2}\left(|0y0\rangle+(-1)^{x}|1\bar{y}0\rangle+|0y^{\prime}1\rangle+(-1)^{x}|1\bar{y^{\prime}}1\rangle\right)_{A_{1}A_{2}..A_{n}ABC}\\
 & = & \left(\alpha\left|a_{1}e_{1}...e_{n-1}\right\rangle +\beta\left|\bar{a_{1}}e_{1}..e_{n-1}\right\rangle \right)\otimes\frac{1}{2}\left(|0y0\rangle+(-1)^{x}|1\bar{y}0\rangle+|0y^{\prime}1\rangle+(-1)^{x}|1\bar{y^{\prime}}1\rangle\right)_{A_{1}A_{2}..A_{n}ABC}\end{array}\]
 Alice measures the separable qubits ($A_{2}A_{3}...A_{n}$) in computational
basis and this correspond to the measurement $M_{n-1}.$ This measurement reduces the state $|\Psi\rangle_{2}$ to
$|S\rangle_{A_{1}AB}|0\rangle_{C}+|T\rangle_{A_{1}AB}|1\rangle_{C}$ where $|S\rangle_{A_{1}AB}$ and
$|T\rangle_{A_{1}AB}$ can be written as \begin{equation}
\begin{array}{lcl}
|S\rangle_{A_{1}AB} & = & \frac{1}{4}\left(\alpha\left|a_{1}0y\right\rangle +(-1)^{x}\alpha\left|\bar{a_{1}}1y\right\rangle +\beta\left|a_{1}0\overline{y}\right\rangle +(-1)^{x}\beta\left|\bar{a_{1}}1\overline{y}\right\rangle \right)_{A_{1}AB}\\
 & = & \frac{1}{2\sqrt{2}}\left(\left(\frac{\left|a_{1}0\right\rangle +(-1)^{x}\left|\bar{a_{1}}1\right\rangle }{\sqrt{2}}\right)_{A_{1}A}\otimes(\alpha\left|y\right\rangle +\beta|\overline{y})_{B}+\left(\frac{\left|a_{1}0\right\rangle -(-1)^{x}\left|\bar{a_{1}}1\right\rangle }{\sqrt{2}}\right)_{A_{1}A}\otimes(\alpha\left|y\right\rangle -\beta\left|\overline{y}\right\rangle )_{B}\right.\\
 & + & \left.\left(\frac{(-1)^{x}\left|a_{1}1\right\rangle +\left|\bar{a_{1}}0\right\rangle }{\sqrt{2}}\right)_{A_{1}A}\otimes(\alpha\left|\overline{y}\right\rangle +\beta|y)_{B}+\left(\frac{(-1)^{x}\left|a_{1}1\right\rangle -\left|\bar{a_{1}}0\right\rangle }{\sqrt{2}}\right)_{A_{1}A}\otimes(\alpha\left|\overline{y}\right\rangle -\beta|y)_{B}\right)\end{array}\label{eq:A-decompose}\end{equation}
 and \begin{equation}
\begin{array}{lcl}
|T\rangle_{A_{1}AB} & = & \frac{1}{4}\left(\alpha\left|a_{1}0y'\right\rangle +(-1)^{x'}\alpha\left|\bar{a_{1}}1y'\right\rangle +\beta\left|a0\overline{y'}\right\rangle +(-1)^{x'}\beta\left|\bar{a_{1}}1\overline{y'}\right\rangle \right)_{A_{1}AB}\\
 & = & \frac{1}{2\sqrt{2}}\left(\left(\frac{\left|a_{1}0\right\rangle +(-1)^{x^{\prime}}\left|\bar{a_{1}}1\right\rangle }{\sqrt{2}}\right)_{A_{1}A}\otimes(\alpha\left|y^{\prime}\right\rangle +\beta|\overline{y^{\prime}})_{B}+\left(\frac{\left|a_{1}0\right\rangle -(-1)^{x^{\prime}}\left|\bar{a_{1}}1\right\rangle }{\sqrt{2}}\right)_{A_{1}A}\otimes(\alpha\left|y^{\prime}\right\rangle -\beta\left|\overline{y^{\prime}}\right\rangle )_{B}\right.\\
 & + & \left.\left(\frac{(-1)^{x^{\prime}}\left|a_{1}1\right\rangle +\left|\bar{a_{1}}0\right\rangle }{\sqrt{2}}\right)_{A_{1}A}\otimes(\alpha\left|\overline{y^{\prime}}\right\rangle +\beta|y^{\prime})_{B}+\left(\frac{(-1)^{x^{\prime}}\left|a_{1}1\right\rangle -\left|\bar{a_{1}}0\right\rangle }{\sqrt{2}}\right)_{A_{1}A}\otimes(\alpha\left|\overline{y^{\prime}}\right\rangle -\beta|y^{\prime})_{B}\right)\end{array}\label{eq:B-decompose}\end{equation}
respectively. From (\ref{eq:channel2}), (\ref{eq:A-decompose}) and (\ref{eq:B-decompose}) one can easily see
that if Alice measures the qubits 1 and 2 using Bell analyzers $M_{BM}$ and Charlie measures his qubit by a
measurement $M_{C}$ in computational basis then the state of Bob reduces to a one qubit state which can be
transformed into the unknown qubit state by applying an unitary transformation. The choice of the particular
unitary operation depends on the outcome of the measurement of Alice ($M_{BM})$ and that of Charlie ($M_{C}$).
Therefore, when Alice and Charlie sent the outcomes of their measurements ($M_{BM}$ and $M_{C}$ ) to Bob using
classical channels then Bob can choose appropriate unitary operations as per the Table 3 to construct
$\alpha|x_{1}\rangle+\beta|\bar{x_{1}}\rangle$. After that Bob creates $n-1$ ancillas
$\left|B_{1}B_{2}..B_{n-1}\right\rangle $, such that $\left|B_{1}B_{2}..B_{n-1}\right\rangle $=
$\left|e_{1}..e_{n-1}\right\rangle $ and follow the last part of the protocol described in section 2 to
reconstruct the unknown state with the help of $n-1$ Cnot operations as shown in Fig. 4.

From Table 3, it is clear that the every member of a family of quantum channels described by
$\frac{(\psi_{i}\left|0\right\rangle +\psi_{j}\left|1\right\rangle )}{\sqrt{2}}$ can be used for multiparty
quantum teleportation. Yang \emph{et al.} \cite{Yang} and Zhang \emph{et al.} \cite{Zhang} did special cases of
the present work. But unfortunately there are some mistakes in the work of Yang \emph{et al.} \cite{Yang}.
Present work provides a generalized perception of the quantum channel and provide more options for
experimentalist to experimentally realize the quantum information splitting. Further the earlier protocols
\cite{Zhang,Yang} were designed for CT of an unknown qubit $\alpha|0\rangle+\beta|1\rangle$ only but here it is
extended for CT of much more general and complex state of the form (\ref{eq:state}).

\begin{table}
\begin{longtable}{|ccccc>{\raggedright}p{1in}|}
\hline \multicolumn{1}{||c|}{{\tiny Quantum Channel}} & \multicolumn{1}{c|}{{\tiny $M_{BM}$}} &
\multicolumn{1}{c|}{{\tiny $M_{C}$}} & \multicolumn{1}{c|}{{\tiny $U$}} & {\tiny $M_{C}$ } & {\tiny
$U_{B}$}\tabularnewline \hline {\tiny $\frac{1}{\sqrt{2}}\left(\psi^{^{+}}\left|0\right\rangle
+\psi^{^{-}}\left|1\right\rangle \right)$ } & {\tiny $\psi_{A_{1}A}^{+}$ } & {\tiny $\left|0\right\rangle _{C}$
} & {\tiny I } & {\tiny $\left|1\right\rangle _{C}$ } & {\tiny Z}\tabularnewline
 & {\tiny $\psi_{A_{1}A}^{-}$ } & {\tiny $\left|0\right\rangle _{C}$ } & {\tiny Z } & {\tiny $\left|1\right\rangle _{C}$ } & {\tiny I}\tabularnewline
 & {\tiny $\phi_{A_{1}A}^{+}$ } & {\tiny $\left|0\right\rangle _{C}$ } & {\tiny X } & {\tiny $\left|1\right\rangle _{C}$ } & {\tiny XZ}\tabularnewline
 & {\tiny $\phi_{A_{1}A}^{-}$ } & {\tiny $\left|0\right\rangle _{C}$ } & {\tiny XZ } & {\tiny $\left|1\right\rangle _{C}$ } & {\tiny X}\tabularnewline
{\tiny $\frac{1}{\sqrt{2}}\left(\psi^{^{+}}\left|0\right\rangle +\phi^{+}\left|1\right\rangle \right)$ } &
{\tiny $\psi_{A_{1}A}^{+}$ } & {\tiny $\left|0\right\rangle _{C}$ } & {\tiny I } & {\tiny $\left|1\right\rangle
_{C}$ } & {\tiny X}\tabularnewline
 & {\tiny $\psi_{A_{1}A}^{-}$ } & {\tiny $\left|0\right\rangle _{C}$ } & {\tiny Z } & {\tiny $\left|1\right\rangle _{C}$ } & {\tiny XZ}\tabularnewline
 & {\tiny $\phi_{A_{1}A}^{+}$ } & {\tiny $\left|0\right\rangle _{C}$ } & {\tiny X } & {\tiny $\left|1\right\rangle _{C}$ } & {\tiny I}\tabularnewline
 & {\tiny $\phi_{A_{1}A}^{-}$ } & {\tiny $\left|0\right\rangle _{C}$ } & {\tiny XZ } & {\tiny $\left|1\right\rangle _{C}$ } & {\tiny Z}\tabularnewline
{\tiny $\frac{1}{\sqrt{2}}\left(\psi^{^{+}}\left|0\right\rangle +\phi^{^{-}}\left|1\right\rangle \right)$ } &
{\tiny $\psi_{A_{1}A}^{+}$ } & {\tiny $\left|0\right\rangle _{C}$ } & {\tiny I } & {\tiny $\left|1\right\rangle
_{C}$ } & {\tiny XZ}\tabularnewline
 & {\tiny $\psi_{A_{1}A}^{-}$ } & {\tiny $\left|0\right\rangle _{C}$ } & {\tiny Z } & {\tiny $\left|1\right\rangle _{C}$ } & {\tiny X}\tabularnewline
 & {\tiny $\phi_{A_{1}A}^{+}$ } & {\tiny $\left|0\right\rangle _{C}$ } & {\tiny X } & {\tiny $\left|1\right\rangle _{C}$ } & {\tiny Z}\tabularnewline
 & {\tiny $\phi_{A_{1}A}^{-}$ } & {\tiny $\left|0\right\rangle _{C}$ } & {\tiny XZ } & {\tiny $\left|1\right\rangle _{C}$ } & {\tiny I}\tabularnewline
{\tiny $\frac{1}{\sqrt{2}}\left(\psi^{-}\left|0\right\rangle +\psi^{+}\left|1\right\rangle \right)$ } & {\tiny
$\psi_{A_{1}A}^{+}$ } & {\tiny $\left|0\right\rangle _{C}$ } & {\tiny Z } & {\tiny $\left|1\right\rangle _{C}$ }
& {\tiny I}\tabularnewline
 & {\tiny $\psi_{A_{1}A}^{-}$ } & {\tiny $\left|0\right\rangle _{C}$ } & {\tiny I } & {\tiny $\left|1\right\rangle _{C}$ } & {\tiny Z}\tabularnewline
 & {\tiny $\phi_{A_{1}A}^{+}$ } & {\tiny $\left|0\right\rangle _{C}$ } & {\tiny XZ } & {\tiny $\left|1\right\rangle _{C}$ } & {\tiny X}\tabularnewline
 & {\tiny $\phi_{A_{1}A}^{-}$ } & {\tiny $\left|0\right\rangle _{C}$ } & {\tiny X } & {\tiny $\left|1\right\rangle _{C}$ } & {\tiny XZ}\tabularnewline
{\tiny $\frac{1}{\sqrt{2}}\left(\psi^{^{-}}\left|0\right\rangle +\phi^{+}\left|1\right\rangle \right)$ } &
{\tiny $\psi_{A_{1}A}^{+}$ } & {\tiny $\left|0\right\rangle _{C}$ } & {\tiny Z } & {\tiny $\left|1\right\rangle
_{C}$ } & {\tiny X}\tabularnewline
 & {\tiny $\psi_{A_{1}A}^{-}$ } & {\tiny $\left|0\right\rangle _{C}$ } & {\tiny I } & {\tiny $\left|1\right\rangle _{C}$ } & {\tiny XZ}\tabularnewline
 & {\tiny $\phi_{A_{1}A}^{+}$ } & {\tiny $\left|0\right\rangle _{C}$ } & {\tiny XZ } & {\tiny $\left|1\right\rangle _{C}$ } & {\tiny I}\tabularnewline
 & {\tiny $\phi_{A_{1}A}^{-}$ } & {\tiny $\left|0\right\rangle _{C}$ } & {\tiny X } & {\tiny $\left|1\right\rangle _{C}$ } & {\tiny Z}\tabularnewline
{\tiny $\frac{1}{\sqrt{2}}\left(\psi^{^{-}}\left|0\right\rangle +\phi^{^{-}}\left|1\right\rangle \right)$ } &
{\tiny $\psi_{A_{1}A}^{+}$ } & {\tiny $\left|0\right\rangle _{C}$ } & {\tiny Z } & {\tiny $\left|1\right\rangle
_{C}$ } & {\tiny XZ}\tabularnewline
 & {\tiny $\psi_{A_{1}A}^{-}$ } & {\tiny $\left|0\right\rangle _{C}$ } & {\tiny I } & {\tiny $\left|1\right\rangle _{C}$ } & {\tiny X}\tabularnewline
 & {\tiny $\phi_{A_{1}A}^{+}$ } & {\tiny $\left|0\right\rangle _{C}$ } & {\tiny XZ } & {\tiny $\left|1\right\rangle _{C}$ } & {\tiny Z}\tabularnewline
 & {\tiny $\phi_{12}^{-}$ } & {\tiny $\left|0\right\rangle _{C}$ } & {\tiny X } & {\tiny $\left|1\right\rangle _{C}$ } & {\tiny I}\tabularnewline
{\tiny $\frac{1}{\sqrt{2}}\left(\phi^{+}\left|0\right\rangle +\phi^{-}\left|1\right\rangle \right)$ } & {\tiny
$\psi_{12}^{+}$ } & {\tiny $\left|0\right\rangle _{C}$ } & {\tiny X } & {\tiny $\left|1\right\rangle _{C}$ } &
{\tiny XZ}\tabularnewline
 & {\tiny $\psi_{A_{1}A}^{-}$ } & {\tiny $\left|0\right\rangle _{C}$ } & {\tiny XZ } & {\tiny $\left|1\right\rangle _{C}$ } & {\tiny X}\tabularnewline
 & {\tiny $\phi_{A_{1}A}^{+}$ } & {\tiny $\left|0\right\rangle _{C}$ } & {\tiny I } & {\tiny $\left|1\right\rangle _{C}$ } & {\tiny Z}\tabularnewline
 & {\tiny $\phi_{A_{1}A}^{-}$ } & {\tiny $\left|0\right\rangle _{C}$ } & {\tiny Z } & {\tiny $\left|1\right\rangle _{C}$ } & {\tiny I}\tabularnewline
{\tiny $\frac{1}{\sqrt{2}}\left(\phi^{^{+}}\left|0\right\rangle +\psi^{+}\left|1\right\rangle \right)$ } &
{\tiny $\psi_{A_{1}A}^{+}$ } & {\tiny $\left|0\right\rangle _{C}$ } & {\tiny X } & {\tiny $\left|1\right\rangle
_{C}$ } & {\tiny I}\tabularnewline
 & {\tiny $\psi_{A_{1}A}^{-}$ } & {\tiny $\left|0\right\rangle _{C}$ } & {\tiny XZ } & {\tiny $\left|1\right\rangle _{C}$ } & {\tiny Z}\tabularnewline
 & {\tiny $\phi_{A_{1}A}^{+}$ } & {\tiny $\left|0\right\rangle _{C}$ } & {\tiny I } & {\tiny $\left|1\right\rangle _{C}$ } & {\tiny X}\tabularnewline
 & {\tiny $\phi_{A_{1}A}^{-}$ } & {\tiny $\left|0\right\rangle _{C}$ } & {\tiny Z } & {\tiny $\left|1\right\rangle _{C}$ } & {\tiny XZ}\tabularnewline
{\tiny $\frac{1}{\sqrt{2}}\left(\phi^{+}\left|0\right\rangle +\psi^{-}\left|1\right\rangle \right)$ } & {\tiny
$\psi_{A_{1}A}^{+}$ } & {\tiny $\left|0\right\rangle _{C}$ } & {\tiny Z } & {\tiny $\left|1\right\rangle _{C}$ }
& {\tiny X}\tabularnewline
 & {\tiny $\psi_{A_{1}A}^{-}$ } & {\tiny $\left|0\right\rangle _{C}$ } & {\tiny I } & {\tiny $\left|1\right\rangle _{C}$ } & {\tiny XZ}\tabularnewline
 & {\tiny $\phi_{A_{1}A}^{+}$ } & {\tiny $\left|0\right\rangle _{C}$ } & {\tiny XZ } & {\tiny $\left|1\right\rangle _{C}$ } & {\tiny I}\tabularnewline
 & {\tiny $\phi_{A_{1}A}^{-}$ } & {\tiny $\left|0\right\rangle _{C}$ } & {\tiny X } & {\tiny $\left|1\right\rangle _{C}$ } & {\tiny Z}\tabularnewline
{\tiny $\frac{1}{\sqrt{2}}\left(\phi^{-}\left|0\right\rangle +\phi^{+}1\right)$ } & {\tiny $\psi_{A_{1}A}^{+}$ }
& {\tiny $\left|0\right\rangle _{C}$ } & {\tiny Z } & {\tiny $\left|1\right\rangle _{C}$ } & {\tiny
XZ}\tabularnewline
 & {\tiny $\psi_{A_{1}A}^{-}$ } & {\tiny $\left|0\right\rangle _{C}$ } & {\tiny I } & {\tiny $\left|1\right\rangle _{C}$ } & {\tiny X}\tabularnewline
 & {\tiny $\phi_{A_{1}A}^{+}$ } & {\tiny $\left|0\right\rangle _{C}$ } & {\tiny X } & {\tiny $\left|1\right\rangle _{C}$ } & {\tiny Z}\tabularnewline
 & {\tiny $\phi_{A_{1}A}^{-}$ } & {\tiny $\left|0\right\rangle _{C}$ } & {\tiny XZ } & {\tiny $\left|1\right\rangle _{C}$ } & {\tiny I}\tabularnewline
{\tiny $\frac{1}{\sqrt{2}}\left(\phi^{-}\left|0\right\rangle +\psi^{+}\left|1\right\rangle \right)$ } & {\tiny
$\psi_{A_{1}A}^{+}$ } & {\tiny $\left|0\right\rangle _{C}$ } & {\tiny XZ } & {\tiny $\left|1\right\rangle _{C}$
} & {\tiny I}\tabularnewline
 & {\tiny $\psi_{A_{1}A}^{-}$ } & {\tiny $\left|0\right\rangle _{C}$ } & {\tiny X } & {\tiny $\left|1\right\rangle _{C}$ } & {\tiny Z}\tabularnewline
 & {\tiny $\phi_{A_{1}A}^{+}$ } & {\tiny $\left|0\right\rangle _{C}$ } & {\tiny Z } & {\tiny $\left|1\right\rangle _{C}$ } & {\tiny X}\tabularnewline
 & {\tiny $\phi_{A_{1}A}^{-}$ } & {\tiny $\left|0\right\rangle _{C}$ } & {\tiny I } & {\tiny $\left|1\right\rangle _{C}$ } & {\tiny XZ}\tabularnewline
{\tiny $\frac{1}{\sqrt{2}}\left(\phi^{-}\left|0\right\rangle +\psi^{-}1\right)$ } & {\tiny $\psi_{A_{1}A}^{+}$ }
& {\tiny $\left|0\right\rangle _{C}$ } & {\tiny X } & {\tiny $\left|1\right\rangle _{C}$ } & {\tiny
Z}\tabularnewline
 & {\tiny $\psi_{A_{1}A}^{-}$ } & {\tiny $\left|0\right\rangle _{C}$ } & {\tiny XZ } & {\tiny $\left|1\right\rangle _{C}$ } & {\tiny I}\tabularnewline
 & {\tiny $\phi_{A_{1}A}^{+}$ } & {\tiny $\left|0\right\rangle _{C}$ } & {\tiny I } & {\tiny $\left|1\right\rangle _{C}$ } & {\tiny XZ}\tabularnewline
 & {\tiny $\phi_{A_{1}A}^{-}$ } & {\tiny $\left|0\right\rangle _{C}$ } & {\tiny Z } & {\tiny $\left|1\right\rangle _{C}$ } & {\tiny X}\tabularnewline
\hline
\end{longtable}

\caption{Controlled teleportation of $\alpha|x\rangle+\beta|\bar{x}\rangle$ is possible by using quantum
channels of the form $\frac{(\psi_{i}\left|0\right\rangle +\psi_{j}\left|1\right\rangle )}{\sqrt{2}}$. Yang
\emph{et al.} {[}\ref{the:Yang-ghz}{]} have used $\frac{1}{\sqrt{2}}\left(\psi^{^{+}}\left|1\right\rangle
+\phi^{+}\left|0\right\rangle \right)$ as their quantum channel and Zhang \emph{et al.} {[}\ref{the:Zhang}{]}
have used $\frac{1}{\sqrt{2}}\left(\psi^{^{+}}\left|0\right\rangle +\phi^{+}\left|1\right\rangle \right)$ as
their quantum channel for controlled teleportation of $\alpha|0\rangle+\beta|1\rangle$. $U_{B}$ denotes the
unitary operation to be applied by Bob after receiving the outcome of $M_{A_{1}A}$ and $M_{C}$.}

\end{table}

\section{Conclusions}

In our protocol we have used maximally entangled states as quantum channel. It is straight forward exercise to
extend these protocols to the situations in which non-maximally entangled states of the similar form are used as
quantum channel. In that case the success probability of the teleportation will not be unity. Thus the
teleportation schemes would become probabilistic. For example, if we use
$(\alpha|\psi_{i}\rangle\left|0\right\rangle +\beta|\psi_{j\neq i}\rangle|1\rangle)$ as the teleportation
channel in place of GHZ-\emph{like} state in Section 4 or $\alpha\left|00\right\rangle
+\beta\left|11\right\rangle $ in place of $\frac{1}{\sqrt{2}}\left(\left|00\right\rangle +\left|11\right\rangle
\right)$ in Section 2 then we will obtain a probabilistic teleportation scheme. Of course the nature of unitary
operations conducted by Bob will be different. Similarly the existing schemes for probabilistic teleportation
can be easily converted to the scheme for perfect teleportation. There are several proposals for perfect and
probabilistic teleportation \cite{Lu,Shi2,Cao,Li}. Our scheme is economical compared to most of these proposals
because our quantum channel consumes only one Bell state. Since we have used only a Bell state for perfect
quantum teleportation of $n$-qubit Bell state and a GHZ for controlled teleportation of $n$-qubit Bell state. It
is obvious that the minimum amount of non-local quantum resource is used here as quantum channel. The same
amount of resource are used by An Ba \cite{baan} in their excellent proposal of teleporting
$\alpha\left|0\right\rangle ^{\otimes n}+\beta\left|1\right\rangle ^{\otimes n}$ using an EPR state. Our
proposal is more general as it can teleport $\alpha\left|x\right\rangle +\beta\left|\bar{x}\right\rangle $ in
general. Further An Ba's protocol may be obtained as a special case of our protocol. As in An Ba case
$\left|e_{1}e_{2}..e_{n-1}\right\rangle $ will always be $|0\rangle^{\otimes n-1}.$ We don't need the
corresponding Cnot gates in encoding circuit in Fig. 1. We don't even need to measure or classically communicate
them. Consequently, Bob always prepares his ancillas in state $\left|B_{1}B_{2}..B_{n-1}\right\rangle
=\left|0\right\rangle ^{\otimes n-1}$. The present protocol is not only more generalized compared to An Ba
protocol \cite{baan} but it is much more simpler too. Further our protocol always require only two measurement
and two classical communications to teleport $\alpha\left|0\right\rangle ^{\otimes n}+\beta\left|1\right\rangle
^{\otimes n}$ but the An Ba protocol always require three measurements and sometime three classical
communications too. Since the quantum gates, the quantum channels and the measurements used in the present
proposals are already experimentally achieved \cite{qt1,qt2,qt3,qt4,qt5,qt6}. We may claim that our protocols
which involve minimum non-local quantum resource is experimentally achievable. The biggest advantage of the
proposed circuits lies in the fact that maintaining a multi-partite entangled quantum channel is costly. In our
case the channel is optimal as far as the number of qubits are concerned. We have proposed two schemes for CT.
One is by using GHZ state as quantum channel; and other is by using GHZ-\emph{like} states as quantum channel.
It is needless to say that the works in \cite{Zhang} and \cite{Yang} are just special cases of our proposal.
Further we would like to note that An Ba generalized the work of Cola \cite{Cola} and consequently the work of
Cola is a special case of An Ba and since An Ba protocol is special case of our protocol so Cola \cite{Cola}
protocol is also a special case of our protocol. Further an improved version of Gorbachev protocol for
teleportation of 2-qubit entangled state is also obtained as a special case here. Another advantage of the
present work lies in the fact that unitary operations to be performed by Bob is always one qubit operation. In
many of the existing protocols Bob needs to implement more complex quantum gates. We end this paper with a
justified expectation that the present proposals and their variants will  find applications in dense coding,
measurement less quantum error correction and secured quantum communication.

\section*{Acknowledgments}

We thank R. Srikanth for some helpful discussions.


\end{document}